\documentclass[%
 aip,
 jmp,%
 amsmath,amssymb,
 reprint,%
]{revtex4-1}

\usepackage{graphicx}
\usepackage{dcolumn}
\usepackage{bm}

\begin{document}


\title{Energy deposition dynamics of femtosecond pulses in water}

\author{Stefano Minardi}
\email{stefano@stefanominardi.eu.}
\homepage{http://stefanominardi.eu.}
\affiliation{Institute of Applied Physics, Friedrich-Schiller-Universit\"{a}t Jena, Max-Wien-Platz 1, 07743 Jena, Germany}

\author{Carles Mili\'an}
\affiliation{Centre de Physique Th{\'e}orique, CNRS, {\'E}cole Polytechnique, F-91128 Palaiseau, France}
 
\author{Donatas Majus}
\affiliation{Department of Quantum Electronics, Vilnius University, Sauletekio 9, bldg. 3, LT-10222 Vilnius, Lithuania}

\author{Amrutha Gopal}
\affiliation{Institute of Optics and Quantum Electronics, Friedrich-Schiller-Universit\"{a}t Jena, Max-Wien-Platz 1, 07743 Jena, Germany}

\author{Gintaras Tamo\v{s}auskas} 
\affiliation{Department of Quantum Electronics, Vilnius University, Sauletekio 9, bldg. 3, LT-10222 Vilnius, Lithuania}

\author{Arnaud~Couairon}
\affiliation{Centre de Physique Th{\'e}orique, CNRS, {\'E}cole Polytechnique, F-91128 Palaiseau, France}

\author{Thomas Pertsch}%
\affiliation{Institute of Applied Physics, Friedrich-Schiller-Universit\"{a}t Jena, Max-Wien-Platz 1, 07743 Jena, Germany}

\author{Audrius Dubietis}
\affiliation{Department of Quantum Electronics, Vilnius University, Sauletekio 9, bldg. 3, LT-10222 Vilnius, Lithuania}

\date{\today}

\begin{abstract}
We exploit inverse Raman scattering and solvated electron absorption to perform 
a quantitative characterization of the energy loss and ionization dynamics in water with tightly focused near-infrared femtosecond pulses. 
A comparison between experimental data and numerical simulations suggests that the ionization energy of water is 8 eV, rather than the commonly used value of 6.5 eV. We also introduce an equation for the Raman gain valid for ultra-short pulses that validates our experimental procedure. 
\end{abstract}

\maketitle

Femtosecond laser pulses tightly focused in dielectric media have a wide range of applications in science and technology. 
Because of their capability to deposit high ionization doses in volumes of a few cubic microns, they can be used to 
induce permanent, microscopic refractive index modification in solid dielectrics, thus enabling three-dimensional integrated optics
\cite{Nolte, Thomson}.
By focusing femtosecond pulses in liquids, it is possible to induce localized chemical reactions such as photo-polymerization on the micro-nano-scale \cite{Vilnius}. 
In aqueous media, such as biological tissues, tightly focused femtosecond laser pulses have been successfully employed for eye surgery \cite{Kammel} and treatment of cancerous cells \cite{PNAS2012}. 
Recent studies show that by tuning the input pulse chirp an effective control on the energy deposition in water is reached\cite{CM1}. Future developments of these applications will benefit from a more advanced control of the energy deposition by means of arbitrarily spatiotemporally tailored laser wavepackets \cite{Razvan14}.
In this context, suitable diagnostic tools for real time analysis of energy deposition dynamics as well as a better understanding of the initial stages of the energy absorption in the dielectric medium are of foremost importance. 

In previous experiments based on quantitative shadowgraphy,
we characterized the propagation of a 120 fs pulse focused with low NA in water \cite{Stem1,Stem2} .
In this configuration, the laser pulse enters a filamentation regime \cite{Arnaud} leaving behind a tenuous, few-mm-long plasma channel which gets solvated on a ps timescale. 
The pulse dynamics (featuring pulse splitting and superluminal pulse formation) was clearly seen in the probe as an absorption feature, which we attributed to the imaginary part of an unspecified cross-phase modulation process (XPM) between pump and probe.

\begin{figure}
\includegraphics[width=8.3 cm]{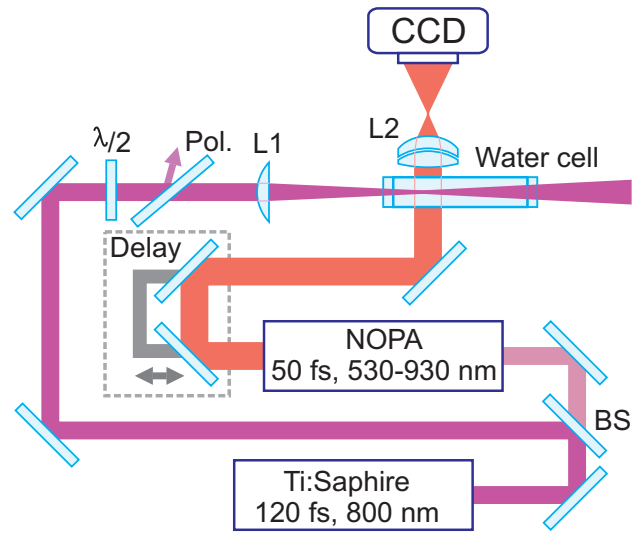}
\caption{\label{setup} The experimental set-up. $\lambda/2$: rotating lambda-half plate. Pol.: polarizer. L1: pump focusing lens (NA$\sim0.08$). L2: microscope objective (NA=0.25).}
\end{figure}

Here we explore the energy deposition dynamics of tightly focused near infrared fs-pulses in water. We provide experimental evidence that the previously reported XPM absorption of the weak probe pulse is in fact inverse Raman scattering (IRS) \cite{PRL13}, allowing us to exploit single shadowgrams to extract the energy of the pump wavepacket and its evolution on microscopic propagation lengths. A crucial assumption behind this procedure is that the absorption of the probe is linear with the energy of the pump. We have verified that this holds by means of a generalized equation for the probe intensity, valid for ultra-short fs-pulses. By combining the extracted pump energy with the peak plasma density estimated from the absorption traces of solvated electrons, we have reconstructed the energy loss dynamics of the pump pulse.
Comparison between the losses estimated from the IRS traces and from the electron solvation traces suggests that the ionization threshold of water is 8 eV, rather than the commonly used value of 6.5 eV. Numerical simulations of the pulse propagation corroborate this finding.  


A scheme of the setup is depicted in Figure \ref{setup}. The 120 fs pulse from the Ti:Sapph laser is split among two channels. One channel (the pump) is focused by lens L1 at $NA=W_{1/e^2}/f\sim0.08$ in the middle of a quartz cuvette filled with purified water ($W_{1/e^2}$ is the waist radius of the Gaussian beam). The energy of the pump beam is regulated by a $\lambda/2$ plate and polarizer to $810\pm5$ nJ, so that a stable, $\sim 200\,\mu$m-long trace of solvated electrons can be excited. Energy of the pump was kept below the threshold for broadband supercontinuum generation and cavitation bubble formation \cite{cavitation}.
The second channel (probe) consists of 50 fs pulses tunable from 530 to 930 nm and delivered by a second-harmonic-pumped non-collinear optical parametric amplifier (NOPA) (Topas White - Light Conversion Ltd.). The probe beam is collimated and illuminates the cuvette perpendicular to the direction of propagation of the pump.  A long working distance microscope equipped with a high-resolution CCD camera is then used to record the transmitted beam profile with a transverse spatial resolution of 2 $\mu$m and a plate scale of 0.5 $\mu$m/pixel. The camera focuses the plane containing the plasma trace so that only transmission measurements are possible with the shadowgram. 

\begin{figure}[t]
\includegraphics[width=8.3 cm]{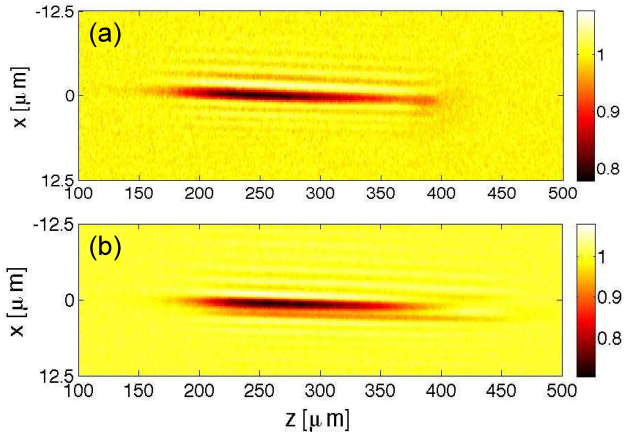}
\caption{\label{frame}(a) Shadowgram at $\lambda_{\mathrm{probe}}=625$ nm after noise removal by local averaging with a 1$\mu$m$\times 1\mu$m stencil. The propagation direction of the pulse is from left to right. An absorption peak at $z=375\,\mu$m is followed by the trace of the solvated electrons ($z\sim250\,\mu$m). (b)  The corresponding shadowgram at $\lambda_{\mathrm{probe}}=800$ nm. Notice that the shadowgram does not show any trace of the pump pulse.
Fringes around the solvated plasma channel are due to the finite numerical aperture of the microscope objective.}
\end{figure}

\begin{figure}
\includegraphics[width=8.2 cm]{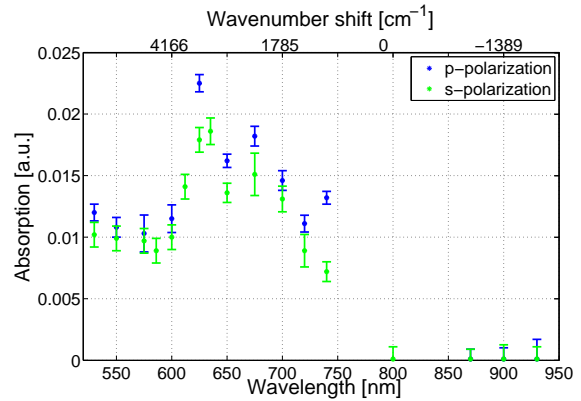}\\
\caption{\label{spectrumIRS} Absorption spectrum of the shadow of the pump pulse. Peak absorption is found at $\lambda_\mathrm{probe}=625$ nm, which corresponds to a detuning of 3500 cm$^{-1}$ from the pump frequency ($\lambda_\mathrm{pump}=800$ nm). }
\end{figure}

Initially, we studied the spectral characteristics of the absorption features appearing in the shadowgrams, by recording a sequence of instant images (taken by scanning the delay of the probe in steps of 50 fs) of the focusing pump pulse for 17 probe wavelengths between 530 and 930 nm. We noticed that for a fixed position in space and probe wavelength shorter than 740 nm, the time profile of the absorption signal was characterized by a sharp peak ($\sim$100-150 fs duration) followed by a slowly growing signal (time constant 500 fs, see Fig. \ref{frame}(a)). In previous experiments with $\lambda_{\mathrm{probe}} = 560$ nm (Ref.\onlinecite{Stem1}), we interpreted this feature as a signature of the pump pulse followed by the buildup of the solvation of a free-electron plasma. 
On the contrary, for probe wavelengths larger than 740 nm, only the solvation curve was observed and no signature of the pump pulse was evident (Fig.\ref{frame}(b) ).
By measuring the peak absorption of the pump pulse feature as a function of the probe wavelength we obtained the spectrum depicted in Fig. \ref{spectrumIRS}.
The spectrum is peaked at $\lambda_{\mathrm{probe}}=$625 nm and its profile is not significantly sensitive to the polarization state of the probe beam. A similar spectral feature has been previously observed in supercontinuum spectra of filaments in water \cite{Deepak} and attributed to IRS \cite{PRL13}.
Indeed, the frequency difference between the probe and the pump corresponds precisely to the central frequency $\nu\sim3500\,$cm$^{-1}$ of the hydroxyl bond stretching  vibrational band, which is $\Delta\nu\approx$ 400 cm$^{-1}$ wide \cite{Walrafen}.

In the limit of nearly monochromatic light, the shadowgraphic signal $F_{\mathrm{shad},t}(x,z)$ at $\lambda_{\mathrm{probe}}=625$ nm due to the IRS can be written as \cite{Rahn1981}:
\begin{equation}
F_{\mathrm{shad},t}(x,z)=F_{\mathrm{probe}}\exp\left[-g\int I_{\mathrm{pump}}(x,y,z,t)dy\right].
\end{equation}
Here $I_{\mathrm{pump}}(x,y,z,t)$ is the intensity of the pump pulse evaluated at the time $t$ (laboratory frame) at which the impulsive, plane wave probe pulse of fluence $F_{\mathrm{probe}}$ crosses the pump pulse. The constant $g$ is the Raman gain which, for the transition at  $\nu\sim3500\,$cm$^{-1}$, takes the value of 0.14 cm/GW (Ref. \onlinecite{Kiraz2009}). The energy of the pump wavepacket can be then easily obtained from the spatially resolved absorbance ($A=-\log_{10} T$, where $T=F_{\mathrm{shad},t}(x,z)/F_{\mathrm{probe}}$)  and then integrating it in the plane $(x,z)$. 



\begin{figure}[t]
\includegraphics[width=8.3 cm]{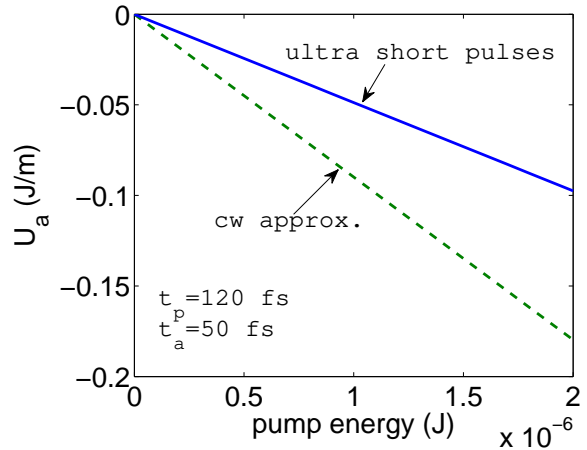}
\caption{\label{reduced_gain} The plot shows the probe absorption calculated as a function of the pump pulse energy. Pump and probe test pulses are taken as spatio-temporal Gaussian packets $\sim\exp(-x_i^2/w_i^2-t^2/t_i^2)$: $w_i=5$ $\mu$m, and $t_{a,p}$ given in figure.}
\end{figure}

Eq. (1) follows from the anti-stokes probe intensity variation under the action of the pump in the case of cw beams: $\partial_yI_\mathrm{probe}=-gI_\mathrm{pump}I_\mathrm{probe}$. We generalized this equation to make it valid for ultra-short pump and probe fields by following the standard procedure\cite{Agrawal} for the case of homogeneous media. Note that because of the negligible role played by the different polarizations in the absorption (see Fig. \ref{spectrumIRS}) we neglected the vectorial effects\cite{vector} in our derivation. By decomposing the real electric field in the pump plus anti-stokes complex fields, ${\mathcal{E}}(x,y,z,t)=\frac{1}{2}[E_\mathrm{pump}e^{-i\omega_\mathrm{pump}t}+E_\mathrm{probe}e^{-i\omega_\mathrm{probe}t}]+c.c.$ and neglecting any linear losses we obtain the linearized equation for the probe intensity, ($I\equiv|E|^2$):
\begin{eqnarray}
 \partial_yI_\mathrm{probe}&=&-[g+\Delta g(E_{\mathrm{probe},\mathrm{pump}})]I_\mathrm{pump}I_\mathrm{probe},\\
g&\equiv& k_0f_R\chi^{(3)}{\mathrm{Im}}\left\{\tilde{h}_R(\Omega)\right\},\\
\Delta{g}&\equiv& k_0f_R\chi^{(3)}{\mathrm{Im}}\left\{\sum_{k=1}^{\infty}c_k(\Omega)\frac{\partial_t^k(E_\mathrm{probe}E_\mathrm{pump}^*)}{E_\mathrm{probe}E_\mathrm{pump}^*}\right\}, \\
c_k&\equiv&\int_{0}^{\infty}dx\frac{(-x)^k}{k!}h_R(x)e^{i\Omega x},
\end{eqnarray}
where $k_0\equiv\omega_\mathrm{probe}/c$, $f_R=0.33$ is the relative nonlinear Raman fraction\cite{note}, $\chi^{(3)}=\frac{8}{3}n_0n_2$ ($n_0=1.33$, $n_2=2.7\times10^{-16}$ cm$^2$/W) is the third order susceptibility, $h_{R}\left(t\right)\equiv\Theta(t)\frac{\tau_{1}^{2}+\tau_{2}^{2}}{\tau_{1}\tau_{2}^{2}}\exp(-t/\tau_2)\sin(t/\tau_1)$ is the Raman response function, ($\tau_1=[2\pi c\nu]^{-1}=1.52$ fs, $\tau_2=2/\Delta\nu[$THz$]=26.5$ fs) and $\tilde h_R(\Omega)=c_0$ its spectrum evaluated at the frequency $\Omega\equiv\omega_\mathrm{probe}-\omega_\mathrm{pump}$. Space time integration of the right hand side of Eq. (2) gives the energy loss per unit length, $U_a$, of the probe pulse. In Fig. \ref{reduced_gain}, we show that $U_a$ vs the energy of the pump pulse at the moment of maximum spatial overlap follows a linear trend, consistently with early theoretical studies\cite{Bloem}, albeit with a lower effective gain as compared to the monochromatic limit: $0<g+\Delta g<g$.


\begin{figure}[t]
\includegraphics[width=8.3 cm]{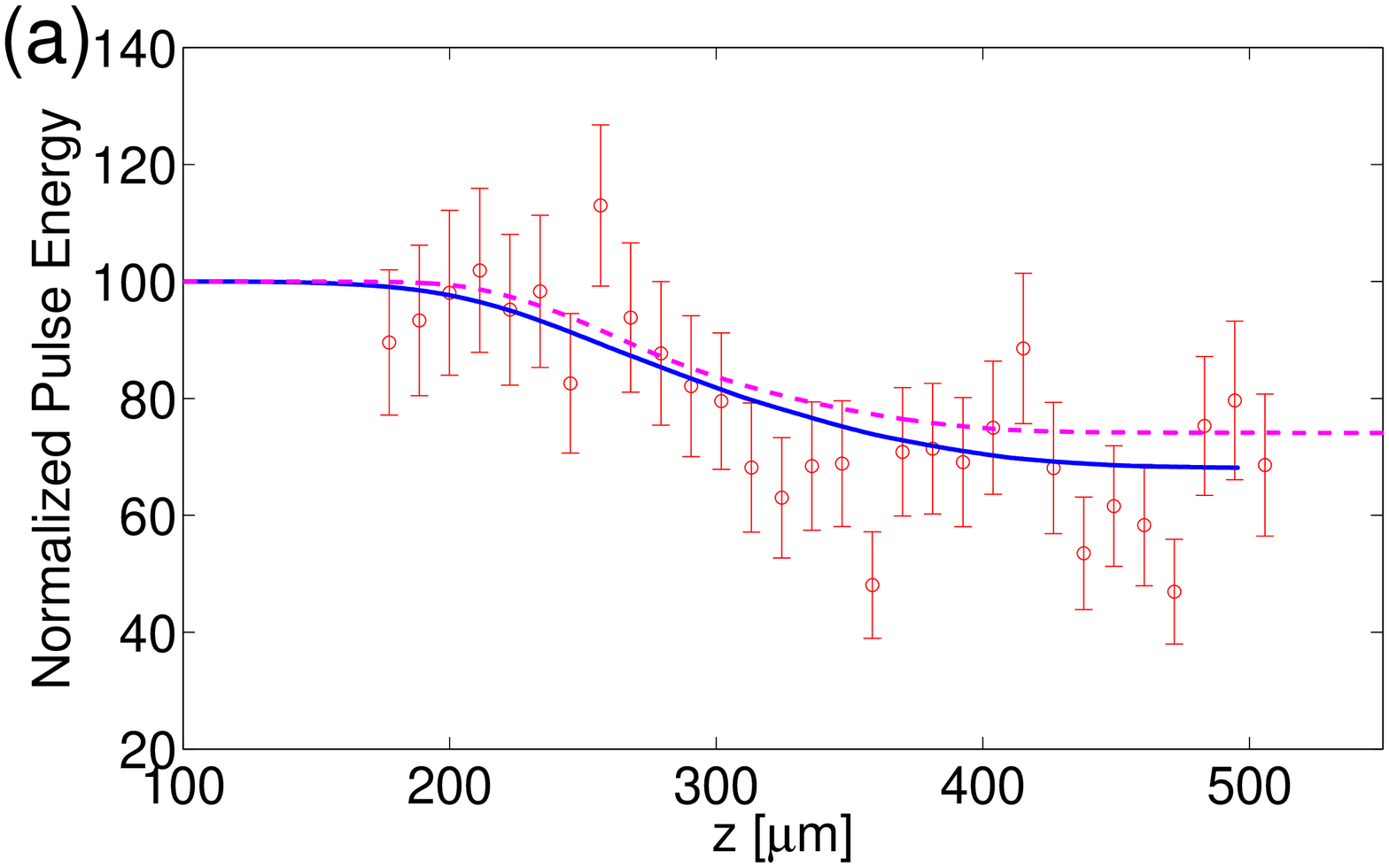}
\includegraphics[width=8.3 cm]{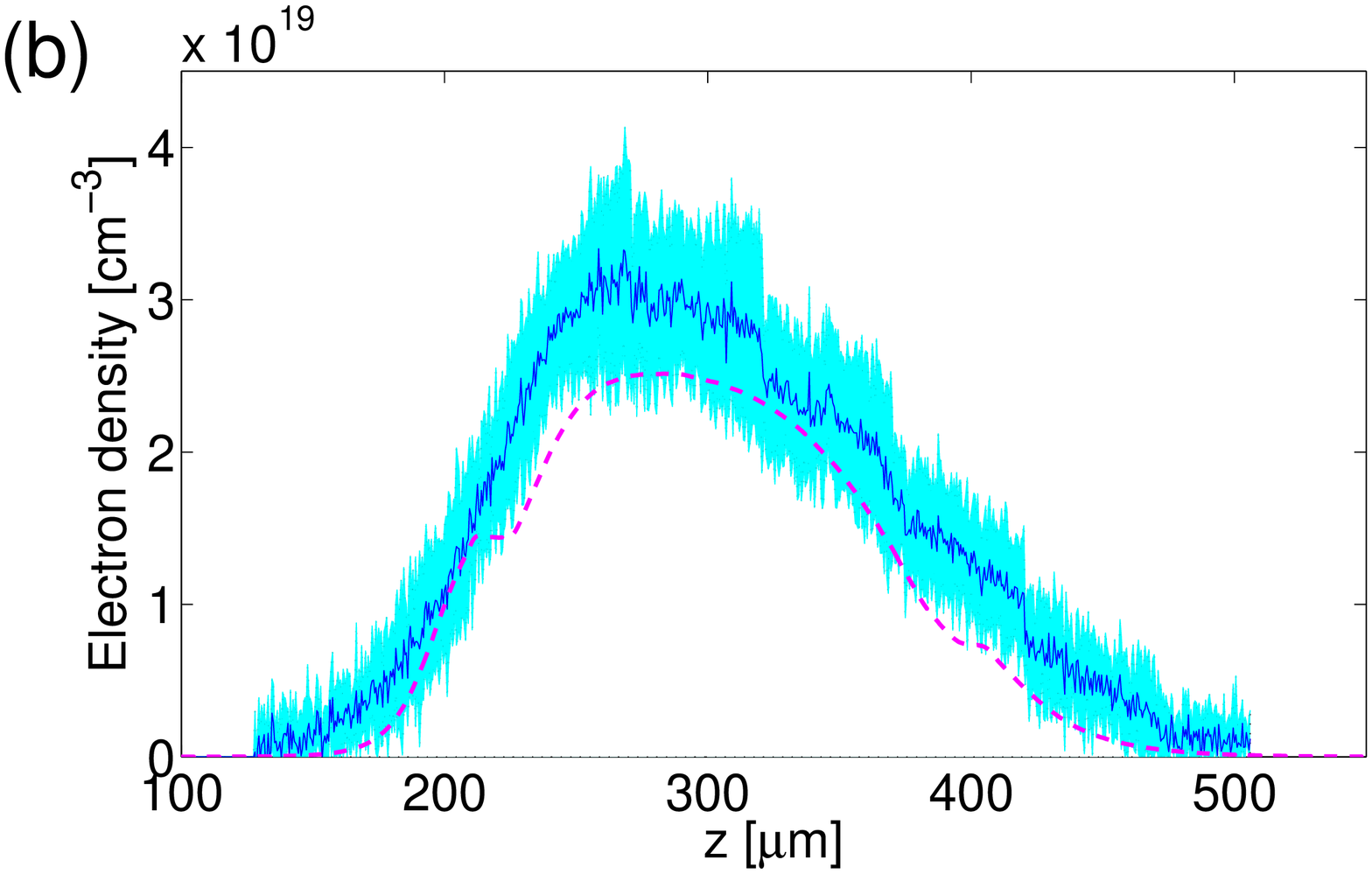}
\caption{\label{plasma} (a) Spatially resolved energy deposition in the water cuvette. Points: estimate of the losses from the analysis of the IRS traces of the pump pulse. Continuous line: losses estimated from the integration of the measured electron density profile. Dashed line: energy losses as from the numerical simulation of the experiment. (b) Measured solvated plasma density. The shade indicates the error bar interval of our estimate. Dashed line: estimated solvated plasma density from the numerical simulation.}
\end{figure}

The above results demonstrate that we can use the shadowgrams to measure \textit{in-situ} the energy of the pump pulse as it propagates through the sample. By integrating the absorbance on a small window centered around the position of the pump pulse peak in each frame of the time delayed shadowgrams, we obtained the plot of Fig. \ref{plasma} (a).  The data show that the pump wavepacket loses about 30\% of its energy within a $200\,\mu$m propagation.

We next verified the trend of the energy loss by analyzing quantitatively the absorption data of the solvated electrons tail. 
By Abel inversion of the absorbance at $\lambda_{\mathrm{probe}}=$625 nm of 5 images taken at a late delay (few hundreds of fs after the pump pulse transit), we obtained the radial distribution of the absorption coefficient\cite{Sun2005}. This was related to a solvated electron density by using the known peak decadic molar extinction coefficient ($\epsilon_{720}=1.85\times 10^4$ cm$^{-1}$mol$^{-1}$dm$^3$, Ref. \onlinecite{Migus87}) suitably scaled to the wavelength of 625 nm from the absorption spectrum of the plasma channel ($\epsilon_{625}=1.32\times 10^4$ cm$^{-1}$mol$^{-1}$dm$^3$), which we measured (data not shown).
The peak solvated plasma density as a function of the propagation length $z$ is depicted in Fig. \ref{plasma}(b). 
The plasma channel had a uniform FWHM diameter of $3.0\pm0.5\,\mu$m.

Numerical simulations were carried out to have more insight in the energy deposition process. To this end we used a model for nonlinear pulse propagation in the presence of plasma. The pulse propagation is described by a Unidirectional Nonlinear Envelope Equation \cite{AC1} including sources accounting for Kerr self-focusing, plasma defocusing and nonlinear losses due to plasma formation and plasma absorption. This equation is solved together with a rate equation for the plasma density, which includes excitation of electrons using the Keldysh ionization rate \cite{Keldysh}, as well as recombination with time $\tau_r = 100$ fs (Ref. \onlinecite{Stem2}). The plasma is modeled as a Drude gas, so that the inverse Bremsstrahlung cross section is proportional to the plasma resistivity regulated by a single parameter $\tau_c=3$ fs, describing the collision time of the plasma. The pulse parameters are fixed (NA=0.072, pulse duration 120 fs and energy 810 nJ). 
To convert the plasma density in the solvated electron density, we considered a scenario in which the quantum yield of the solvated electrons depends on the total energy of the multi-photon transition responsible for the excitation of a free electron \cite{Bartels2000}.
In the simulation, we used a ionization potential of 8 eV, which is considered as the correct ionization potential of water, according to recent literature \cite{Bernas1998}. This ionization potential corresponds to a total multi-photon transition of 9.24 eV, and a solvation yield of $44\%$ (compare with Ref. \onlinecite{Bartels2000}).
Note that the standard ionization potential value of 6.5 eV (Ref. \onlinecite{Kennedy1995}), corresponding to a total multi-photon transition energy of 7.7 eV and a fraction of only $4\%$ of solvated electrons\cite{Bartels2000}, did not allow us to reproduce quantitatively the solvated electron density and absorption simultaneously. 
In contrast, with a gap of 8 eV, we were able to fit precisely both
the measured solvated electron curve and the energy losses (see Fig. \ref{plasma}).  By post-processing the measurement data of the plasma distribution and assuming that a 6-photon process is required to excite
one molecule, we obtained a loss curve which nicely fits the loss data obtained from the measurements using IRS shadows, provided we further assume that only $50\%$ of the produced excitations form a solvated electron (see Fig. \ref{plasma}(a)).

Concluding, we showed that by tuning the probe wavelength to the Raman anti-stokes  wavelength, intense optical pulses propagating in condensed matter dielectrics can be visualized as transmission dips in shadowgrams. Simultaneously, the single shadowgram can provide information on the plasma density, if some form of color-center can be generated after the relaxation of the electrons. In our water samples we were able to combine the information about the spatiotemporal dynamics the pump pulse with a precise estimate of the deposited energy.
Our numerical investigation combined with experimental findings pointed out that the relevant ionization process for a $\lambda_0=800$ nm pump is a 6-photon-absorption, confirming that the ionization potential of water is $U_g\sim8$ eV.

S.M. and A.G. acknowledge financial support from European Commission Seventh Framework Programme project LASERLAB-EUROPE access, Grant No. 228334. D.M., G.T. and A.D. acknowledge financial support from the European Social Fund under the Global Grant measure (Grant No. VP1-3.1- \v{S}MM-07-K-03-001).
A.C. and C.M. acknowledge financial support from the French DGA.


\begin{thebibliography}{99}
\bibitem{note} This value is chosen as to achieve consistency in between the known value $g=0.14$ cm/GW and Eq. (3).
\end{thebibliography}
\end{document}